\documentclass[journal,twoside,web]{IEEEtran}
\usepackage{cite}
\usepackage{dutchcal}
\usepackage{amsmath,amssymb,amsfonts}
\usepackage{algorithmic}
\usepackage{hyperref}       
\usepackage{url}            
\usepackage{booktabs}       
\usepackage{amsfonts}       
\usepackage{nicefrac} 
\usepackage{microtype}      
\usepackage{lipsum}
\usepackage{graphicx}
\usepackage{textcomp}
\usepackage{float}
\usepackage{makecell}
\usepackage{multirow}
\usepackage{color}
\usepackage{diagbox}
\usepackage{hyperref}[colorlinks,linkcolor=black]
\usepackage{todonotes}
\usepackage{color,soul}
\usepackage{gensymb}
\def\etal{\emph{et al.}}
\def\ie{\emph{i.e.}}
\def\eg{\emph{e.g.}}

\begin{document}
\title{TBI-GAN: An Adversarial Learning Approach for Data Synthesis on Traumatic Brain Segmentation}
\author{Xiangyu Zhao, Di Zang, Sheng Wang, Zhenrong Shen, Kai Xuan, Zeyu Wei, Zhe Wang, Ruizhe Zheng, Xuehai Wu, Zheren Li, Qian Wang, Zengxin Qi, and Lichi Zhang
\thanks{Corresponding authors: Z. Qi and L. Zhang (e-mail: qizengxin@huashan.org.cn, lichizhang@sjtu.edu.cn).}
\thanks{X. Zhao and D. Zang contributed equally to this work.}
\thanks{X. Zhao, S. Wang, Z. Shen, K. Xuan, Z. Li and L. Zhang are with School of Biomedical Engineering, Shanghai Jiao Tong University, Shanghai, 200030, China (e-mails: \{xiangyu.zhao, wsheng, zhenrongshen, kaixuan, lizheren, lichizhang\}@sjtu.edu.cn).}
\thanks{D. Zang, Z. Wei, Z. Wang, R. Zheng, X. Wu, and Z. Qi are with Department of Neurosurgery, Huashan Hospital, Shanghai Medical College, Fudan University, also with National Center for Neurological Disorders, also with Shanghai Key Laboratory of Brain Function and Restoration and Neural Regeneration, and also with State Key Laboratory of Medical Neurobiology and MOE Frontiers Center for Brain Science, School of Basic Medical Sciences and Institutes of Brain Science, Fudan University (e-mails: \{dzang16, 18301050181, 20211220121\}@fudan.edu.cn, \{ruizhedoctor, wuxuehai2013\}@163.com, qizengxin@huashan.org.cn).}
\thanks{Q. Wang is with School of Biomedical Engineering, ShanghaiTech University, Shanghai, 201210, China (e-mail: wangqian2@shanghaitech.edu.cn).}
}
\maketitle
\bibliographystyle{ieeetr}
\begin{abstract}
Brain network analysis for traumatic brain injury (TBI) patients is critical for its consciousness level assessment and prognosis evaluation, which requires the segmentation of certain consciousness-related brain regions. However, it is difficult to construct a TBI segmentation model as manually annotated MR scans of TBI patients are hard to collect. Data augmentation techniques can be applied to alleviate the issue of data scarcity. However, conventional data augmentation strategies such as spatial and intensity transformation are unable to mimic the deformation and lesions in traumatic brains, which limits the performance of the subsequent segmentation task. To address these issues, we propose a novel medical image inpainting model named TBI-GAN to synthesize TBI MR scans with paired brain label maps. The main strength of our TBI-GAN method is that it can generate TBI images and corresponding label maps simultaneously, which has not been achieved in the previous inpainting methods for medical images. We first generate the inpainted image under the guidance of edge information following a coarse-to-fine manner, and then the synthesized intensity image is used as the prior for label inpainting. Furthermore, we introduce a registration-based template augmentation pipeline to increase the diversity of the synthesized image pairs and enhance the capacity of data augmentation. Experimental results show that the proposed TBI-GAN method can produce sufficient synthesized TBI images with high quality and valid label maps, which can greatly improve the 2D and 3D traumatic brain segmentation performance compared with the alternatives. 
\end{abstract}
\section{Introduction}
\label{sec:introduction}
Traumatic brain injury (TBI) is sudden and severe damage that typically alters the brain structures and functions of many regions\cite{menon2010position}. It is considered one of the leading causes of death and disability worldwide with heavy socio-economic consequences \cite{maas2017traumatic}, which usually affects the consciousness level and even leads to disorders of consciousness. Magnetic resonance imaging (MRI) has been widely applied to examine the consciousness and prognosis status of TBI patients in clinical practice, which involves the patient's brain abnormality assessment and brain network analysis \cite{qin2015different, huang2014self}. Specifically, our previous studies \cite{wu2018white, huo2020neuroimage} have identified 17 brain regions that are significantly correlated to the consciousness state of TBI patients. Thus, identification and segmentation of these brain regions are considered a prerequisite for conducting the brain network analysis and the subsequent MRI-based consciousness level estimation, which can provide a more objective quantitative assessment compared with the conventional manner. However, manual annotating of these regions from MRI scans is generally impractical, therefore it is highly demanded to design an automatic segmentation method for TBI images.

\begin{figure}[ht]
	\centering
	\includegraphics[width=0.48\textwidth]{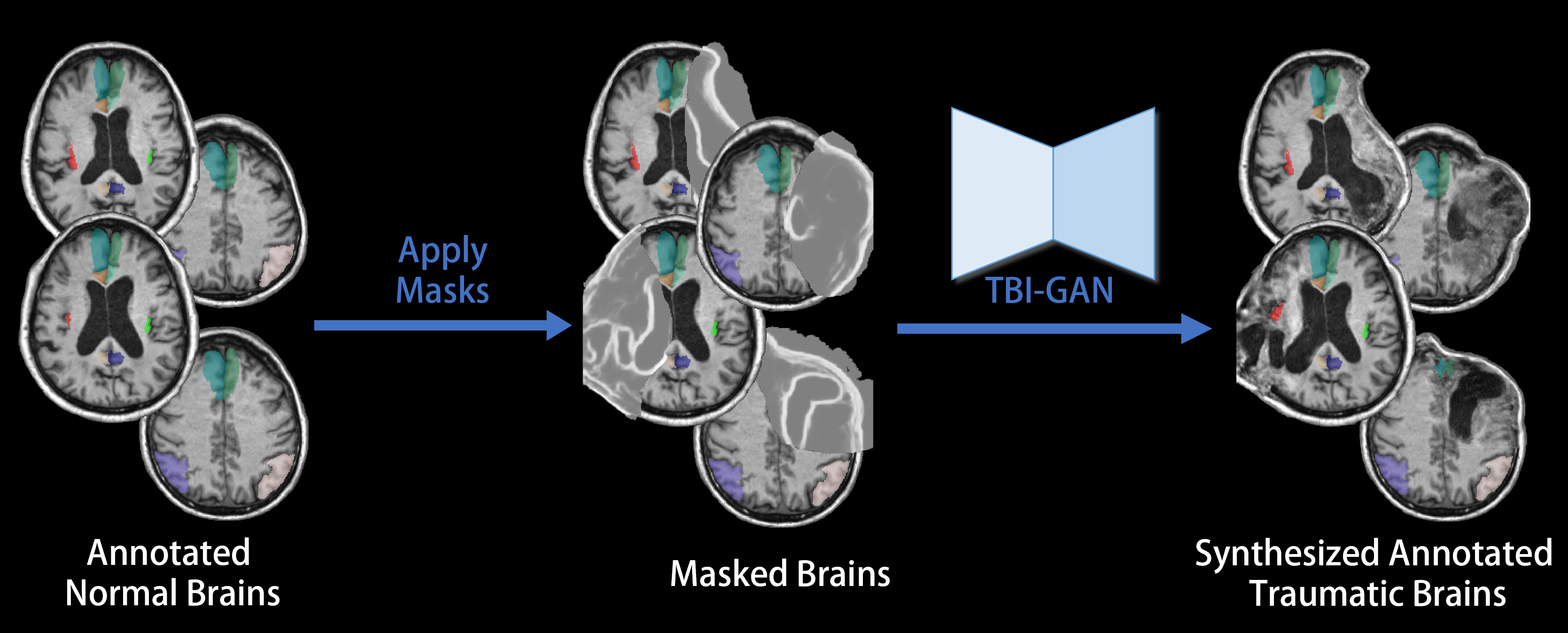} 
	\caption{The workflow of the proposed TBI-GAN. The scheme is drawn in 2D in favor of comprehension.}
	\label{fig:scheme}
\end{figure}

Although many deep-learning-based algorithms have proved to be effective in brain segmentation \cite{huo20193d, ramzan2020volumetric, ren2020robust, qiao2021robust}, current practices are typically trained in a fully-supervised manner, which requires a large amount of annotated data. Due to the intricate anatomical structure of traumatic brain segmentation, the cost of preparation for the training data, especially the manual labeling is quite significant. In this way, data augmentation can be considered an effective way to resolve the issue of lacking training samples. However, despite the popularity of common data augmentation recipes \cite{simard2003best} such as spatial transformation (\eg, affine transform, non-rigid deformation) and intensity transformation (\eg, bright and contrast transform), these techniques are somehow restricted in traumatic brain segmentation. Conventional data augmentation techniques are unable to simulate the anatomical structure of traumatic brains with large deformation and lesion erosion, which limits their effectiveness in producing valid synthesized data and improving the segmentation performance.

Recently, researchers have utilized generative adversarial networks (GANs) to synthesize paired images and label maps for augmentation in medical segmentation tasks \cite{cao2020improving, mok2018learning, platscher2022image}. These methods are typically implemented in an image translation manner, \ie, using the input of semantic label maps to synthesize the corresponding images. Despite the success of these methods, the major limitation of image translation methods for data augmentation is that they need to provide numerous label maps with variations to generate their synthesized images. Since the label maps are also impractical to be collected in application, and it is also difficult to synthesize and augment high-quality label maps which are suitable for generating realistic images, image translation is infeasible to be applied in our case.


Image inpainting \cite{qin2021image} is used to generate the missing content under a masked area, and has been widely applied in image editing. Compared with image translation, image inpainting does not require a large set of sophisticated semantic label maps, which become popular in the data augmentation of classification and object detection \cite{sogancioglu2018chest, shen2021nodule}. However, previous image inpainting methods simply edit the image itself, and cannot provide a corresponding label map of the image, which limits its application in the data augmentation for segmentation tasks. Also, as many traumatic areas are large, these image inpainting methods fail to generate satisfying results in a large masked area. In addition, it remains challenging to ensure the diversity of image inpainting to augment the data for the subsequent tasks.

To address these challenges, here we propose a novel image inpainting network named as TBI-GAN, which can synthesize paired TBI MR images and corresponding brain region label maps given a normal brain template and a lesion mask. The workflow of the proposed TBI-GAN is shown in Fig. \ref{fig:scheme}. Compared with traditional image inpainting methods, our TBI-GAN is able to inpaint both the grayscale image and the corresponding label maps, which can be utilized for the data augmentation of segmentation tasks. Also, we design to inpaint the traumatic area in a coarse-to-fine manner, aiming at producing high-quality inpainting in the lesion masks. Furthermore, we propose to enhance the capacity of the data augmentation by mimicking the diversity of human brains in the population, which proves to be successful in the subsequent segmentation tasks.


The contributions of this work consist of three aspects: 1) We propose a novel network architecture that utilizes image inpainting to generate TBI MR scans and corresponding label maps simultaneously, which has not been achieved in the previous image inpainting methods. 2) We introduce a coarse-to-fine synthesis strategy under the guidance of edge information in the images, which can be utilized to generate high-quality textures in the synthesized TBI images, and the synthesized images are used as the prior to the inpainting of the label maps. 3) We design a registration-based template augmentation pipeline, which can simulate the variance of human brains in the population and improve the diversity of brain trauma, enhancing the capacity of the data augmentation. The proposed TBI-GAN yields state-of-the-art data augmentation performance in both 2D and 3D traumatic brain segmentation tasks.
\begin{figure*}[ht]
	\centering
	\includegraphics[width=0.9\textwidth]{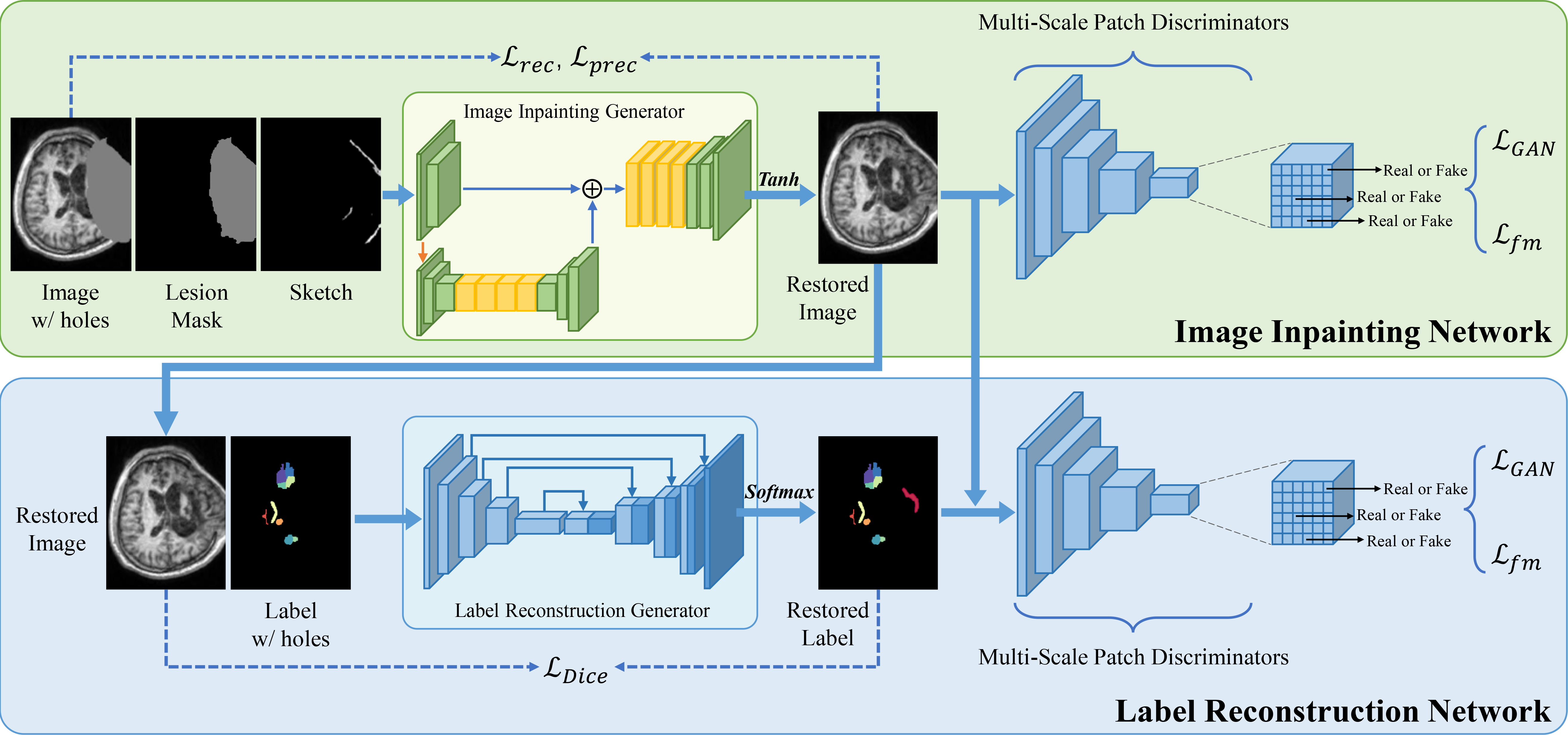} 
	\caption{The overall framework of the proposed TBI-GAN. The framework is drawn in 2D in favor of comprehension.}
	\label{fig:network}
\end{figure*}

\section{Related Works}
\subsection{GAN-based Data Augmentation}
Apart from the conventional data augmentation methods \cite{simard2003best, zhang2017mixup, yun2019cutmix}, GAN methods have also been commonly utilized in the augmentation of medical segmentation problems. Compared with conventional data augmentation practice, GAN-based methods can synthesize realistic-looking images and boost further the performance of the subsequent segmentation task. These algorithms typically feature an image-to-image architecture such as Pix2Pix \cite{isola2017image} or CycleGAN \cite{zhu2017unpaired} to produce realistic synthesized images based on the semantic label of anatomical structures. Thus, the generated images are paired with the semantic label maps to join the training dataset. For example, Shin \etal \cite{shin2018medical} proposed a method to generate synthetic abnormal MRI images with brain tumors from two publicly available brain MR datasets. Mok \etal \cite{mok2018learning} used a coarse-to-fine GAN based on Pix2PixHD \cite{wang2018high} to synthesize brain MR images with tumors from label maps to improve the segmentation performance in brain tumor segmentation. Cao \etal \cite{cao2020improving} proposed a GAN-based augmentation framework to generate multi-modality data pairs on PET and CT to improve the performance of multi-modality segmentation tasks. However, as addressed in Section \ref{sec:introduction} these methods require a large number of label maps to generate their corresponding images, which is infeasible for data augmentation as it is also difficult to collect the label maps in the actual scenario.

\subsection{Image Inpainting}
Image inpainting techniques are widely utilized to fix the defects in images. With the fast development of deep learning, GAN-based image inpainting methods have become dominant in this field. For instance, Liu \etal \cite{liu2018image} proposed the partial convolution operator to replace the vanilla convolution operator in the U-Net to restore the image with the missing region. Yu \etal \cite{yu2018generative} proposed a contextual attention operation to explicitly utilize surrounding image features as references and reconstruct the distorted image. Nazeri \etal \cite{nazeri2019edgeconnect} proposed a two-stage adversarial model EdgeConnect to perform image inpainting under the guidance of image edges. Song \etal \cite{song2018spg} proposed a two-stage inpainting network named SPG-Net to perform image inpainting based on semantic labels. Yu \etal \cite{yu2019free} proposed a gated convolution operation that generalizes partial convolution by introducing a soft gating strategy to solve free-form image inpainting tasks. These image inpainting algorithms are designed for image restoration only and cannot produce label maps directly.

\subsection{Brain Segmentation Based on Deep Learning}
Convolutional neural networks (CNNs) have been widely adopted for brain region segmentation tasks. Ronneberger \etal \cite{ronneberger2015u} proposed U-Net for medical image segmentation, which integrates both low-level and high-level visual representations and has become a baseline algorithm in many medical segmentation tasks. Several works based on U-Net have been applied to brain segmentation. For example, Huo \etal \cite{huo20193d} proposed a spatially localized atlas network tiles method to distribute multiple independent 3D fully convolutional networks for high-resolution whole-brain segmentation. Ramzan \etal \cite{ramzan2020volumetric} introduced 3D CNNs with residual connections and dilated convolution operations to learn the segmentation from brain images efficiently. Such algorithms have been designed to segment normal human brains and can suffer from performance drops when applied to traumatic brains. To solve this problem, Ren \etal \cite{ren2020robust} proposed a hard and soft attention module to be integrated into a segmentation network and achieved brain segmentation in hydrocephalus patients. Qiao \etal \cite{qiao2021robust} developed a robust brain segmentation algorithm via spatial guidance to segment hydrocephalus brains. Despite these recent advances in brain segmentation, these algorithms are fully-supervised which usually requires large amounts of manually annotated training data for constructing the segmentation model, and may suffer from performance deterioration when the data amount is limited.

\section{Methods}
\label{sec:methods}
In this section, we present the details of the proposed TBI-GAN architecture with the image synthesis pipeline. Our proposed TBI-GAN is able to generate multiple traumatic brains with brain region labels based on one brain template. First, we provide an overview of the architecture of the proposed TBI-GAN. Then, we provide the details about the modules in the proposed network, including the image inpainting generator and label reconstruction generator accordingly. We also introduce the registration-based template augmentation pipeline, which can generate multiple normal-looking brains based on one brain template and improve the capacity of the data augmentation.
\subsection{Overview of Network Architecture}
\label{sec:overview}
The overall framework of TBI-GAN is presented in Fig. \ref{fig:network}, which consists of two cascaded networks: 1) image inpainting network aiming at generating trauma lesions inside the mask, and 2) label reconstruction network which generates the corresponding label maps based on the inpainted brain images. Both networks follow the adversarial model configuration in  \cite{goodfellow2014generative}, \ie, each consisting of a pair of generator and discriminator. We use multi-scale discriminators to combine both global and local information, and hence achieve better generation quality. Denote $G_1$ and $D_1$ as the generator/discriminator pair for the image inpainting network, and $G_2$ and $D_2$ for the generator/discriminator pair for the label reconstruction network, respectively.

\subsection{Image Inpainting Network}
\label{sec:IIN}
\begin{figure*}[ht]
	\centering
	\includegraphics[width=0.85\textwidth]{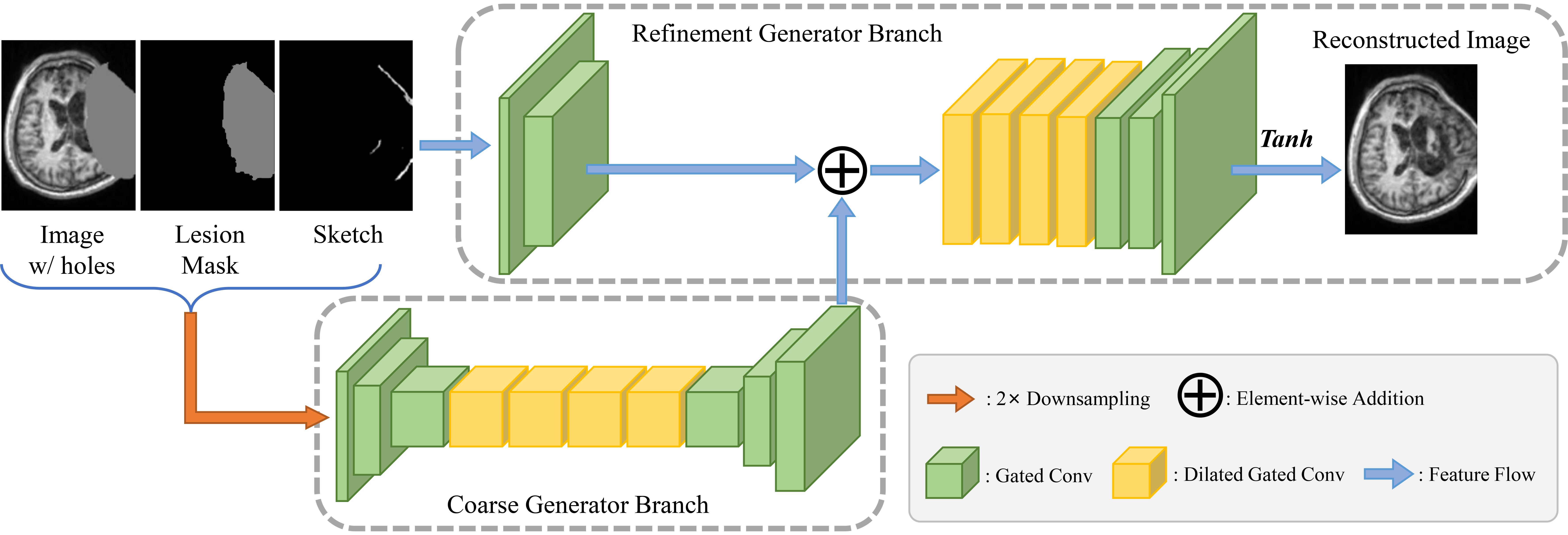} 
	\caption{The network architecture of the proposed image inpainting generator. The image inpainting generator follows a coarse-to-fine manner to synthesize realistic TBI MR scans.}
	\label{fig:image}
\end{figure*}
The image inpainting network is trained to restore the masked traumatic brain images. Let $\mathbf{I}$ denote the ground truth TBI MR scans, the traumatic area is masked by the binary mask $\mathbf{M}$. Then we acquire the masked image $\mathbf{I} \odot (1-\mathbf{M})$, where $\odot$ is the Hadamard product. We also use the HED detector \cite{xie2015holistically} to acquire the sketch of edges slice by slice in the axial direction, which are generally the edges of skulls and ventricles. Since the injuries in the masked areas can be variant, the sketch channel can help facilitate the inpainting of skulls and ventricles. It should be noted that only the sketch inside the masked area, denoted as $\mathbf{S}$, is utilized during training. The image inpainting network is trained to predict the restored TBI MR scans as
\begin{equation}
\mathbf{\tilde{I}}=G_1(\mathbf{I} \odot (1-\mathbf{M}), \mathbf{M}, \mathbf{S}),
\end{equation}
where $\mathbf{\tilde{I}}$ denotes the restored MR image.

\subsubsection{Module Architecture}
The architecture of the generator in the image inpainting network follows a coarse-to-fine manner to generate high-quality MR scans. The detailed network architecture is shown in Fig. \ref{fig:image}. The generator consists of two branches, including a coarse generator branch to produce coarsely predicted feature maps, and a refinement generator branch to generate the final high-quality restored images. Both the branches consist of three components, namely, a gated convolution \cite{yu2019free} downsampling block, a set of dilated gated convolutions, and a gated convolution upsampling block, and the upsampling is achieved by linear interpolation. Note that the inputs are fed into the two branches simultaneously as follows: In the coarse generator branch, the inputs are downsampled and go through convolutional downsampling, dilated bottleneck, and convolutional upsampling to obtain coarse-predicted feature maps. In the refinement generator branch, the inputs are firstly downsampled by convolutional layers. Then the downsampled feature map is added with the output from the coarse generator branch. Finally, the combined feature maps go through a series of dilated convolutions and convolutional upsampling to obtain the refined inpainting results.

\subsubsection{Loss Functions}
We use $\mathbf{I}$ and $\mathbf{\tilde{I}}$ conditioned on $\mathbf{M}$ and $\mathbf{S}$ as inputs to the multi-scale discriminator, which differentiates real or fake MR images. The image inpainting network is trained with the combination of least squares adversarial loss \cite{mao2017least}, $l_1$ reconstruction loss, perceptual loss \cite{johnson2016perceptual, gatys2015texture, gatys2016image} and feature matching loss \cite{wang2018high}.

We also utilize the objective functions described in the least squares generative adversarial networks (LSGAN) \cite{mao2017least} to train the network. The least squares adversarial loss minimizes the Pearson Chi-square divergence during training. Compared with the vanilla GANs loss, least squares adversarial loss encourages better image quality and stabilizes the training. The generator adversarial loss $\mathcal{L}_{G_1}$ and discriminator adversarial loss $\mathcal{L}_{D_1}$ are calculated as follows:
\begin{equation}
\mathcal{L}_{G_1} = \sum^{3}_{k=1} \mathbb{E}_{\mathbf{\tilde{I}} \sim p_{\mathbf{\tilde{I}}}} \ (D_{1k}(\mathbf{\tilde{I}} | \mathbf{M}, \mathbf{S}) - 1)^2,
\end{equation}
\begin{equation}
    \mathcal{L}_{D_1} \! = \! \sum^{3}_{k=1} (\mathbb{E}_{\mathbf{I} \sim p_{\mathbf{I}}} (D_{1k}(\mathbf{I} | \mathbf{M}, \mathbf{S}) - 1)^2
    \!+\! \mathbb{E}_{\mathbf{\tilde{I}} \sim p_{\mathbf{\tilde{I}}}} D_{1k} (\mathbf{\tilde{I}} | \mathbf{M}, \mathbf{S})^2),
\end{equation}
where $D_{1k}$ represents the $k$-th discriminator in the multi-scale discriminators.

We calculate the per-pixel error between the real images and the synthesized images to encourage the correct reconstruction of image structures. The loss is calculated on the whole image rather than the masked area to encourage intensity consistency in the restored image. The $l_1$ reconstruction loss $\mathcal{L}_{rec}$ is defined as
\begin{equation}
\mathcal{L}_{rec} = \mathbb{E}_{\mathbf{I} \sim p_{\mathbf{I}}} \left[ \frac{1}{N} \parallel \mathbf{I} - \mathbf{\tilde{I}} \parallel_1 \right],
\end{equation}
where $N$ denotes the number of elements in the image.

Perceptual loss measures the distance between the synthesized images and real images in the feature space, which is typically extracted by a pretrained encoder. Perceptual loss $\mathcal{L}_{prec}$ is defined as
\begin{equation}
\mathcal{L}_{prec} = \mathbb{E}_{\mathbf{I} \sim p_{\mathbf{I}}} \left[ \sum_{i=1}^{L} \frac{1}{N_i} \parallel \phi_i(\mathbf{I}) - \phi_i(\mathbf{\tilde{I}}) \parallel_1 \right],
\end{equation}
where $\phi$ denotes a VGG-19\cite{simonyan2014very} feature extractor pretrained on ImageNet, $L$ denotes the number of its layers, and $N_i$ denotes the number of elements in the $i$ th activation map.

Feature matching loss compares the feature representations of the synthesized images and real images in the intermediate layers of the discriminator. We introduce feature matching loss to stabilize the training process and accelerate convergence. The feature matching loss $\mathcal{L}_{fm}$ is defined as
\begin{equation}
\mathcal{L}_{fm} \!=\! \mathbb{E}_{\mathbf{I} \sim p_{\mathbf{I}}} \left[ \sum^{3}_{k=1}\sum_{i=1}^{L} \frac{1}{N_i} \!\parallel\! D_{1k}^{(i)}(\mathbf{I} | \mathbf{M}, \mathbf{S}) \!-\! D_{1k}^{(i)}(\mathbf{\tilde{I}} | \mathbf{M}, \mathbf{S}) \!\parallel_1\! \right], \\
\end{equation}
where $L$ denotes the number of layers in the discriminator, and $N_i$ denotes the number of elements in the $i$ th activation map.

In addition, we utilize spectral normalization (SN) \cite{miyato2018spectral} in the discriminator to further stabilize the training, which scales down the weight in the network layers by their largest singular values to encourage 1-Lipschitz continuity.

The final loss function of the image inpainting network is the linear combination of the loss functions above: 
\begin{equation}
\mathcal{L}_{1} = \lambda_{G_1}\mathcal{L}_{G_1} + \lambda_{D_1}\mathcal{L}_{D_1} + \lambda_{rec}\mathcal{L}_{rec} + \lambda_{prec}\mathcal{L}_{prec} + \lambda_{fm}\mathcal{L}_{fm}.
\end{equation}
In the experiments we set $\lambda_{G_1} = 1.0$, $\lambda_{D_1} = 1.0$, $\lambda_{rec} = 1.0$, $\lambda_{prec} = 0.05$ and $\lambda_{fm} = 3.0$.

\subsection{Label Reconstruction Network}
\label{sec:LLN}
The label reconstruction network is trained to restore the brain region labels based on the inpainted images generated by the image inpainting network. Let $\mathbf{A}$ denote the ground truth label map, the label maps inside the traumatic area are also masked by the binary mask $\mathbf{M}$. Thus we obtain the label with holes $\mathbf{A} \odot (1-\mathbf{M})$. The label reconstruction network is trained to generate the label maps inside the masked area, written as:
\begin{equation}
\mathbf{\tilde{A}}=G_2(\mathbf{A} \odot (1-\mathbf{M}) | \mathbf{\tilde{I}}),
\end{equation}
where $\mathbf{\tilde{A}}$ denotes the restored label map.
\subsubsection{Module Architecture}
We adopt the popular U-Net architecture \cite{ronneberger2015u} as the basic architecture of the proposed label reconstruction generator. Compared with the encoder-decoder architecture used in the image inpainting generator, U-Net-based architecture combines both high-level and low-level feature maps and can recover more fine-grained structures in the image. Since the brain region label maps are typically spatially sparse and fine-grained, we adopt U-Net architecture to improve the accuracy of label inpainting. Also, as brain region maps contain almost no structural information, we use the restored image as the conditional input to facilitate the learning process. Such configuration encourages the joint learning of the inpainting of both the images and the labels. 
\subsubsection{Loss Functions}
Similar to the image inpainting generator, we also utilize the least squares adversarial loss to train the adversarial network. Unlike the discriminator used in the image inpainting network, which only recognizes whether the image is real or fake, the discriminator used in the label inpainting network actually predicts whether the paired images and labels are real or fake under the condition of the binary mask. The generator adversarial loss $\mathcal{L}_{G_2}$ and discriminator adversarial loss $\mathcal{L}_{D_2}$ are computed as follows:
\begin{equation}
\mathcal{L}_{G_2} = \sum^3_{k=1} \mathbb{E}_{\mathbf{\tilde{A}} \sim p_{\mathbf{\tilde{A}}}} \ (D_{2k}(\mathbf{\tilde{A}} | \mathbf{\tilde{I}}, \mathbf{M})-1)^2,
\end{equation}
\begin{equation}
    \mathcal{L}_{D_2} \!=\! \sum^3_{k=1} (\mathbb{E}_{\mathbf{A} \sim p_{\mathbf{A}}} (D_{2k}(\mathbf{A} | \mathbf{I}, \mathbf{M})-1)^2 \\
    \!+\! \mathbb{E}_{\mathbf{\tilde{A}} \sim p_{\mathbf{\tilde{A}}}} D_{2k} (\mathbf{\tilde{A}} | \mathbf{\tilde{I}}, \mathbf{M})^2),
\end{equation}
where $D_{2k}$ represents the $k$-th discriminator in the multi-scale discriminators.

Dice loss proposed in \cite{milletari2016v} has been widely used in medical segmentation tasks. Here we utilize Dice loss to facilitate the label inpainting procedure. Similarly, we calculate the Dice loss $\mathcal{L}_{Dice}$ on the whole label map rather than the masked region to ensure anatomical consistency in the restored label maps:
\begin{equation}
\mathcal{L}_{Dice} = 1 - \frac{2| \mathbf{A} \cap \mathbf{\tilde{A}}|}{|\mathbf{A}| + |\mathbf{\tilde{A}}|}
\end{equation}

Also, feature matching loss $\mathcal{L}_{fm}$ is introduced during the training for better convergence:
\begin{equation}
\mathcal{L}_{fm} \!=\! \mathbb{E}_{\mathbf{A} \sim p_{\mathbf{A}}} \!\left[ \sum^{3}_{k=1}\sum_{i=1}^{L} \frac{1}{N_i} \!\parallel\! D_{2k}^{(i)}(\mathbf{A} | \mathbf{I}, \mathbf{M}) \!-\! D_{2k}^{(i)}(\mathbf{\tilde{A}} | \mathbf{\tilde{I}}, \mathbf{M}) \!\parallel_1\! \right]\!
\end{equation}

The final loss function of the label reconstruction network is the linear combination of the loss functions above. 
\begin{equation}
\mathcal{L}_{2} = \lambda_{G_2}\mathcal{L}_{G_2} + \lambda_{D_2}\mathcal{L}_{D_2} + \lambda_{Dice}\mathcal{L}_{Dice} + \lambda_{fm}\mathcal{L}_{fm}
\end{equation}
In the experiments we set $\lambda_{G_2} = 1.0$, $\lambda_{D_2} = 1.0$, $\lambda_{Dice} = 1.0$ and $\lambda_{fm} = 3.0$.

\subsection{Registration-based Template Augmentation}
\begin{figure}[ht]
	\centering
	\includegraphics[width=0.48\textwidth]{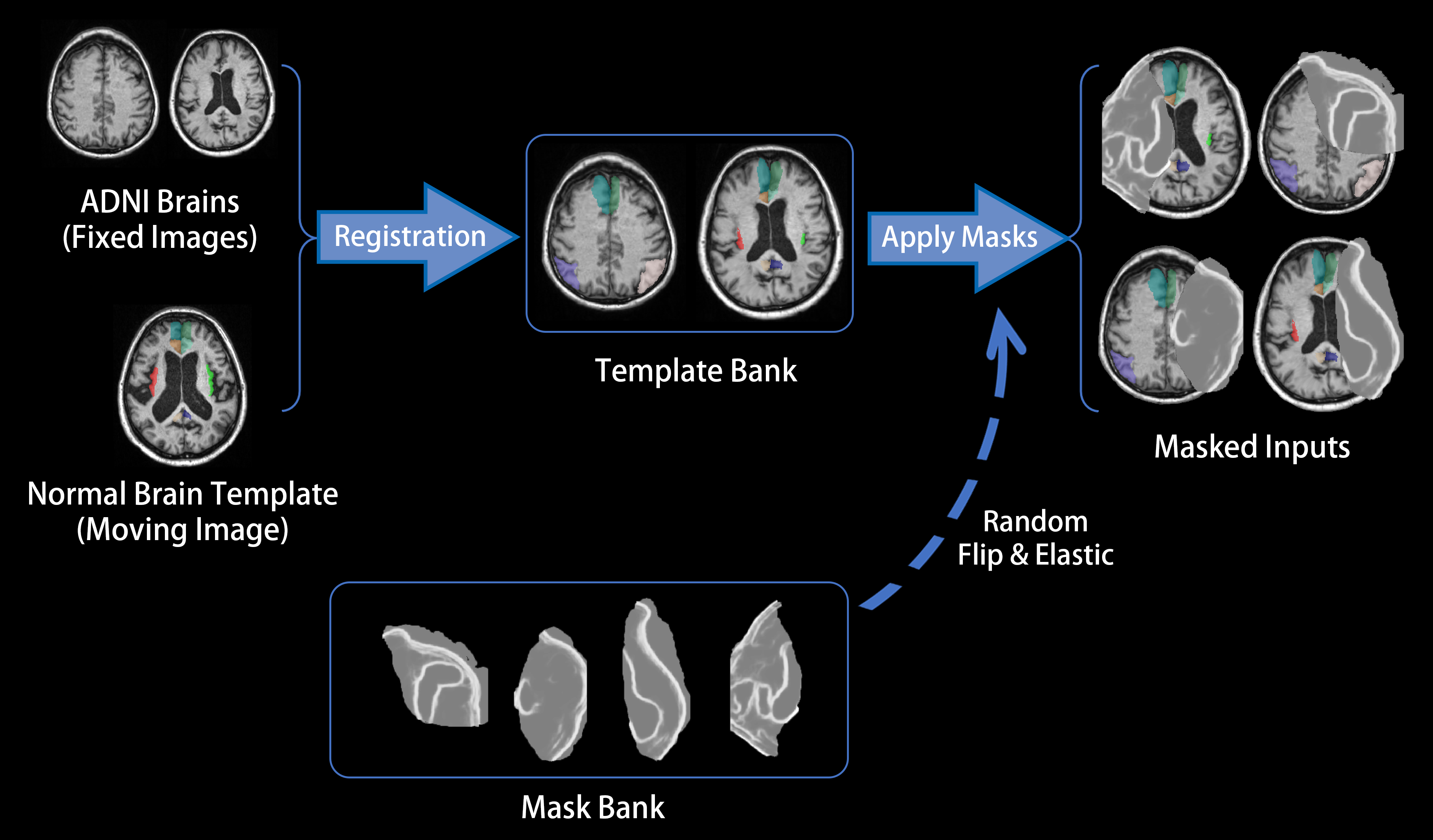} 
	\caption{The overview of the task-optimized inference pipeline. The masked inputs are fed into the TBI-GAN for the synthesis of annotated TBI scans.}
	\label{fig:inference}
\end{figure}

As previously introduced, our TBI-GAN is based on one normal brain template with the to-be-segmented labels. To further increase the diversity of the synthesized image and labels, we design a novel template augmentation procedure aiming at simulating the natural variance of human brains in the population. The overall pipeline is shown in Fig. \ref{fig:inference}. 

The template augmentation pipeline is designed to realize the diversity in both the normal regions and the traumatic areas in the synthesized TBI MR scans. In order to simulate the variance of the human brain population, inspired by \cite{zhao2019data}, we utilize image registration methods to generate multiple normal brain MR images based on one normal brain template. Note that different from \cite{zhao2019data}, we further modify the VoxelMorph \cite{balakrishnan2019voxelmorph} registration framework by cascading multiple registration networks and introducing bending energy loss \cite{rueckert1999nonrigid} during training. Compared with the vanilla VoxelMorph registration network, our modified version achieves better registration accuracy and more realistic deformation fields. The normal brain template has been registered to the unlabeled normal brain MR scans to obtain a set of deformation fields.
Then we apply the deformation fields on both the MR image and the label map of the template to acquire the registered images and their corresponding labels, which builds up a "template bank" for the inpainting procedure.

In addition, we build up a "mask bank" by sampling the masks and sketches from the training set. The sampled masks and sketches from the mask bank are disturbed by random elastic transformation and flipping to simulate the variance of the brain trauma. Next, the transformed mask and sketch are applied to the normal brain image sampled from the brain bank, which is then fed into the TBI-GAN for the synthesis of TBI images and labels. As discontinuity of the sketch and the edges in the normal brain may occur, opening and closing operations are utilized to ensure the consistency of the edges in the normal brain image and the sampled sketch. After synthesis, the generated images and labels are examined to ensure their quality. Failure cases are excluded from the subsequent data augmentation task.
\section{Experiments}
\subsection{Materials}
In this study, we recruited 42 TBI patients and obtained their T1 scans at Huashan Hospital, Fudan University. Informed consent was obtained from all patients for the use of their information, medical records, and MRI data. All MR images were acquired on a 3T Siemens MR scanner. The scanning parameters were: 176 slices, 1 mm slice thickness, no inter-slice gap, repetition time (TR) = 2300 ms, echo time (TE) = 2.98 ms, inversion time = 900 ms, non-interpolated voxel size = 1$\times$1$\times$1 mm$^3$, flip angle = 9$\degree$, and field of view = 240 $\times$ 256 mm$^2$. All MR scans are linearly registered with the MNI152 T1-weighted template using FSL \cite{smith2004advances}. Histogram equalization has also been applied to ensure the stability of intensity contrast from the collected MRIs.

After collecting the TBI dataset, the 17 consciousness-related ROIs according to \cite{wu2018white} have been delineated manually, which are Insula-R (IR), Insula-L (IL), Thalamus-R (TR), Thalamus-L (TL), internal Capsule-R-Ant (ICRA), internal Capsule-R-Post (ICRP), internal Capsule-L-Ant (ICLA), internal Capsule-L-Post (ICLP), Cingulate-R-Ant (CRA), Cingulate-R-Post (CRP), Cingulate-L-Ant (CLA), Cingulate-L-Post (CLP), Medial Prefrontal Cortex-R (MCR), Medial Prefrontal Cortex-L (MCL), Inferior Parietal Lobule-L (IPL), Inferior Parietal Lobule-R (IPR) and Brainstem (B).  

To obtain the normal brain template, we also collect a subject with only ventriculomegaly in imaging diagnosis, as the recruited TBI patients generally suffer from ventriculomegaly simultaneously. The normal brain template was acquired and preprocessed using the same configurations as the TBI dataset, along with the manual annotation procedure. We also find and collect 35 subjects with ventriculomegaly from the ADNI dataset as the fixed images in the "template bank". Similar preprocessing procedures including linear registration to MNI152 and histogram equalization have also been applied to the ADNI MR scans.

\begin{figure*}
	\centering
	\includegraphics[width=0.9\textwidth]{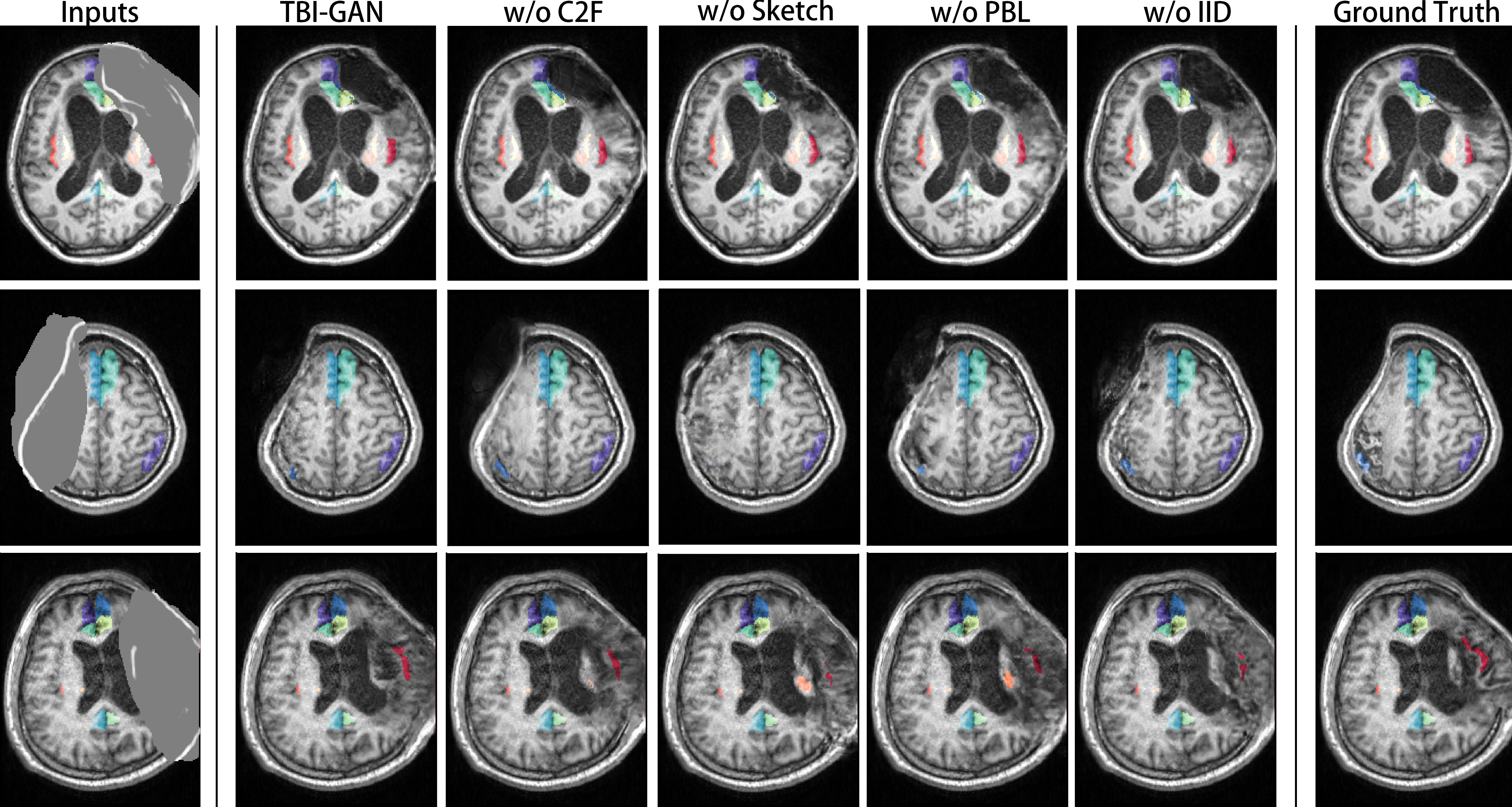} 
	\caption{Qualitative analysis for the proposed modules in TBI-GAN. The input image, lesion mask and sketch are stacked together for a better view.}
	\label{fig:ablation}
\end{figure*}

\subsection{Evaluation of Image Synthesis}
\subsubsection{Implementation Details}
The proposed TBI-GAN network is implemented using the PyTorch 1.9.0 framework on Ubuntu servers. We utilize distributed data parallel on 4 NVIDIA RTX 3090 GPUs to accelerate the training of TBI-GAN. PyTorch native mixed-precision training is utilized to reduce the video memory occupation and further improve the training speed. We have extracted 5329 axial slices with traumatic regions from the collected TBI MR scans. The proposed TBI-GAN is implemented in a 2.5D manner, that it combines the neighboring 3 slices as the actual input, corresponding to the three channels in 2D images. During inference, for each synthesized slice stack, only the middle slice is used as the actual output except for the first/last slice in the 3D volume, where the output is the corresponding slice in the stack \cite{xiang2018deep}. Such training configuration encourages spatial consistency in the axial direction, which can produce realistic 3D volume outputs and benefit subsequent segmentation tasks.

The original spatial size of the axial slices is $182 \times 218$. We resize these slices into $176 \times 224$ with spatial cropping or padding. The model is optimized with the Adam optimizer \cite{kingma2014adam} with $\beta_1 = 0.5$ and $\beta_2 = 0.999$. The initial learning rate of the generators and discriminators are set to $5 \times 10^{-4}$ and $2 \times 10^{-4}$ respectively, and is slowly reduced following the cosine annealing strategy. The batch size is set to 48. We use instance normalization \cite{ulyanov2016instance} as the normalization layers in both the generators and discriminators, which has proved to be effective in many image synthesis tasks. All the upsampling operations in the generators are implemented by bi-linear interpolation rather than transposed convolutions to avoid potential "checkerboard artifacts". Image synthesis evaluation is conducted with a 5-fold cross-validation. It is worth noting that the data is strictly split according to subjects to avoid potential data leakage if slices of the same subject simultaneously exist in the training and validation set.

\subsubsection{Ablation Analysis of the Proposed Modules}
\label{modules}
Here we conduct the ablation study to demonstrate the effectiveness of the modules introduced in the framework. We measure the quality of the synthesized images using the following metrics: 1) peak signal-to-noise ratio (PSNR); 2) structural similarity index measure (SSIM); 3) $\mathcal{l}_1$ error; and 4) $\mathcal{l}_2$ error. The results are shown in Table \ref{table:PSNR}. The visualization of the inpainting results on the validation set is shown in Fig. \ref{fig:ablation}.
\begin{table}
	\centering
	\caption{Ablation study with quantitative analysis for the proposed modules in TBI-GAN.}
	\label{table:PSNR}
	\renewcommand\arraystretch{1.0}
	\setlength{\tabcolsep}{2.2mm}{
		\begin{tabular}{c|cccc}
			\toprule
			 Models & PSNR (dB) & SSIM (\%) & $\mathcal{l}_1$ error & $\mathcal{l}_2$ error  \\
			\midrule
			TBI-GAN & \textbf{26.42} & \textbf{87.54} & \textbf{4.94} & \textbf{2.00} \\
			\midrule
			w/o C2F & 22.75 & 72.02 & 9.30 & 4.13 \\
			w/o Sketch & 23.74 & 85.89 & 6.12 & 3.31 \\
			w/o PBL & 24.12 & 80.84 & 6.78 & 2.67\\
			w/o IID & 25.09 & 86.87 & 5.10 & 2.21\\
			\bottomrule
	\end{tabular}}
\end{table}
\paragraph{Effects of Coarse-to-Fine Generation Strategy (C2F)}
We first conduct experiments to compare the proposed TBI-GAN to the same model, but without the coarse-to-fine (C2F) architecture in the image inpainting generator. In Fig. \ref{fig:ablation}, we show the generated images and labels produced by the compared models. We observe that the coarse-to-fine architecture in the image inpainting generator improves the generation quality significantly. The model without such architecture produces images with fewer details but more "fish scale" artifacts in most cases. Also, observed from Table \ref{table:PSNR}, we can see that the coarse-to-fine strategy used in our network contributes the most to the image quality.
\paragraph{Effects of Sketch Channel}
The proposed TBI-GAN uses the sketch channel with edge information to guide the inpainting, especially for the restoration of the skulls. We conduct a comparative experiment to validate the effectiveness of the sketch channel in the proposed TBI-GAN. In Fig. \ref{fig:ablation}, we observe that the synthesized images generated by the model without the sketch channel are rather blurred in the regions of skulls compared with the ones generated by the full model. Also, the model tends to predict a projecting skull even though the real trauma leads to a depressed skull (see the second row). The experimental results show that the sketch channel can guide the inpainting procedure effectively.
\paragraph{Effects of Perceptual-Based Loss (PBL)}
The perceptual loss and feature matching loss are calculated based on image perceptual features rather than the image itself, which improves the quality of the synthesized images and stabilizes the training process. We conduct a comparative experiment to validate the effectiveness of these perceptual-based losses (PBL) in the proposed TBI-GAN. In Fig. \ref{fig:ablation}, the synthesized images generated by the model without perceptual-based losses lack details compared with the ones generated by the full model (see the last raw). In addition, certain noises can be observed in the black areas (see the first and the second raw).
\paragraph{Effects of Image Inpainting Discriminators (IID)}
The proposed TBI-GAN consists of two complete generative adversarial networks, \ie, the cascade of a generator and a discriminator. We conduct an experiment to validate the effectiveness of the discriminators in the image inpainting network. In Fig. \ref{fig:ablation}, the results produced by the model without the image inpainting discriminators (IID) seem to be good at the first glance. However, compared with the full model, the images produced by the model without the image inpainting discriminators are slightly blurred (see the last row), and may have obvious artifacts in some cases (see the first and the second raw).
\begin{table*}
	\centering
	\caption{Quantitative analysis of 2D traumatic brain segmentation with different augmentation techniques. The best two results are shown in \textcolor{red}{red} and \textcolor{blue}{blue} fonts, respectively.}
	\label{table:2dquantitative}
	\renewcommand\arraystretch{0.95}
	\setlength{\tabcolsep}{1.6mm}{
		\begin{tabular}{cccccccccc}
			\toprule
			  & IR & IL & TR & TL & ICRA & ICRP & ICLA & ICLP & CRA  \\
			\midrule
			Baseline & 0.521$\pm$0.287 & 0.521$\pm$0.259 & 0.601$\pm$0.274 & 0.638$\pm$0.218 & \textcolor{blue}{\textbf{0.570$\pm$0.218}} & 0.439$\pm$0.212 & 0.568$\pm$0.220 & 0.504$\pm$0.205 & 0.476$\pm$0.238 \\
			Spatial & 0.548$\pm$0.268 & 0.556$\pm$0.238 & 0.628$\pm$0.263 & 0.679$\pm$0.209 & 0.544$\pm$0.224 & 0.464$\pm$0.214 & 0.570$\pm$0.210 & 0.539$\pm$0.190 & \textcolor{blue}{\textbf{0.517$\pm$0.232}} \\
			Intensity & 0.527$\pm$0.279 & 0.528$\pm$0.245 & 0.581$\pm$0.282 & 0.624$\pm$0.246 & 0.551$\pm$0.223 & 0.449$\pm$0.218 & 0.548$\pm$0.232 & 0.483$\pm$0.209 & 0.465$\pm$0.252 \\
			Composed & \textcolor{blue}{\textbf{0.550$\pm$0.247}} & \textcolor{blue}{\textbf{0.572$\pm$0.228}} & \textcolor{red}{\textbf{0.644$\pm$0.254}} & \textcolor{blue}{\textbf{0.689$\pm$0.202}} & 0.553$\pm$0.204 & \textcolor{red}{\textbf{0.491$\pm$0.192}} & \textcolor{blue}{\textbf{0.573$\pm$0.193}} & \textcolor{blue}{\textbf{0.541$\pm$0.182}} & \textcolor{red}{\textbf{0.519$\pm$0.227}} \\
			MixUp \cite{zhang2017mixup} & 0.511$\pm$0.309 & 0.518$\pm$0.259 & 0.585$\pm$0.288 & 0.653$\pm$0.215 & 0.552$\pm$0.229 & 0.439$\pm$0.220 & 0.552$\pm$0.227 & 0.492$\pm$0.199 & 0.467$\pm$0.238 \\
			CutMix \cite{yun2019cutmix} & 0.513$\pm$0.297 & 0.530$\pm$0.251 & 0.619$\pm$0.265 & 0.642$\pm$0.232 & 0.561$\pm$0.211 & \textcolor{blue}{\textbf{0.471$\pm$0.199}} & 0.566$\pm$0.216 & 0.515$\pm$0.212 & 0.467$\pm$0.237 \\
			\midrule
			TBI-GAN & \textcolor{red}{\textbf{0.614$\pm$0.234}} & \textcolor{red}{\textbf{0.584$\pm$0.227}} & \textcolor{blue}{\textbf{0.635$\pm$0.269}} & \textcolor{red}{\textbf{0.697$\pm$0.205}} & \textcolor{red}{\textbf{0.605$\pm$0.218}} & 0.438$\pm$0.239 & \textcolor{red}{\textbf{0.588$\pm$0.217}} & \textcolor{red}{\textbf{0.548$\pm$0.200}} & 0.502$\pm$0.198 \\
			\toprule
			  & CRP & CLA & CLP & MCR & MCL & IPL & IPR & B & average  \\
			\midrule
			Baseline & 0.496$\pm$0.145 & 0.479$\pm$0.198 & 0.527$\pm$0.149 & 0.606$\pm$0.197 & 0.604$\pm$0.183 & 0.386$\pm$0.248 & 0.429$\pm$0.227 & 0.869$\pm$0.133 & 0.543$\pm$0.241 \\
			Spatial & \textcolor{blue}{\textbf{0.530$\pm$0.143}} & \textcolor{red}{\textbf{0.528$\pm$0.166}} & 0.545$\pm$0.150 & \textcolor{blue}{\textbf{0.632$\pm$0.154}} & \textcolor{blue}{\textbf{0.632$\pm$0.142}} & 0.433$\pm$0.260 & \textcolor{red}{\textbf{0.499$\pm$0.235}} & \textcolor{blue}{\textbf{0.883$\pm$0.104}} & 0.572$\pm$0.228 \\
			Intensity & 0.510$\pm$0.133 & 0.465$\pm$0.213 & 0.526$\pm$0.147 & 0.615$\pm$0.189 & 0.611$\pm$0.169 & 0.400$\pm$0.266 & 0.408$\pm$0.233 & 0.875$\pm$0.097 & 0.539$\pm$0.244 \\
			Composed & 0.525$\pm$0.146 & \textcolor{blue}{\textbf{0.522$\pm$0.177}} & \textcolor{blue}{\textbf{0.549$\pm$0.155}} & 0.615$\pm$0.183 & 0.621$\pm$0.160 & \textcolor{blue}{\textbf{0.436$\pm$0.261}} & \textcolor{blue}{\textbf{0.494$\pm$0.234}} & 0.879$\pm$0.117 & \textcolor{blue}{\textbf{0.575$\pm$0.223}} \\
			MixUp \cite{zhang2017mixup} & 0.510$\pm$0.140 & 0.465$\pm$0.206 & 0.527$\pm$0.150 & 0.596$\pm$0.210 & 0.608$\pm$0.179 & 0.403$\pm$0.264 & 0.435$\pm$0.207 & 0.879$\pm$0.092 & 0.541$\pm$0.244 \\
			CutMix \cite{yun2019cutmix} & 0.499$\pm$0.145 & 0.455$\pm$0.212 & 0.521$\pm$0.143 & 0.603$\pm$0.204 & 0.610$\pm$0.162 & 0.417$\pm$0.262 & 0.417$\pm$0.238 & 0.867$\pm$0.109 & 0.545$\pm$0.240  \\
			\midrule
			TBI-GAN & \textcolor{red}{\textbf{0.553$\pm$0.131}} & 0.503$\pm$0.181 & \textcolor{red}{\textbf{0.584$\pm$0.117}} & \textcolor{red}{\textbf{0.651$\pm$0.115}} & \textcolor{red}{\textbf{0.643$\pm$0.133}} & \textcolor{red}{\textbf{0.472$\pm$0.227}} & 0.481$\pm$0.262 & \textcolor{red}{\textbf{0.889$\pm$0.111}} & \textcolor{red}{\textbf{0.588$\pm$0.224}} \\
			\bottomrule
	\end{tabular}}
\end{table*}
\begin{table*}
	\centering
	\caption{Quantitative analysis of 3D traumatic brain segmentation with different augmentation techniques. The best two results are shown in \textcolor{red}{red} and \textcolor{blue}{blue} fonts, respectively.}
	\label{table:3dquantitative}
	\renewcommand\arraystretch{0.95}
	\setlength{\tabcolsep}{1.6mm}{
		\begin{tabular}{cccccccccc}
			\toprule
			  & IR & IL & TR & TL & ICRA & ICRP & ICLA & ICLP & CRA  \\
			\midrule
			Baseline & 0.584$\pm$0.258 & 0.592$\pm$0.227 & 0.694$\pm$0.237 & 0.733$\pm$0.175 & 0.610$\pm$0.181 & 0.486$\pm$0.229 & \textcolor{blue}{\textbf{0.629$\pm$0.198}} & 0.570$\pm$0.184 & 0.539$\pm$0.188 \\
			Spatial & 0.569$\pm$0.255 & 0.592$\pm$0.223 & 0.707$\pm$0.211 & 0.732$\pm$0.188 & 0.620$\pm$0.170 & 0.504$\pm$0.226 & 0.625$\pm$0.184 & 0.583$\pm$0.178 & 0.546$\pm$0.209 \\
			Intensity & 0.568$\pm$0.271 & 0.594$\pm$0.211 & 0.687$\pm$0.231 & 0.727$\pm$0.181 & \textcolor{blue}{\textbf{0.635$\pm$0.173}} & 0.492$\pm$0.232 & 0.617$\pm$0.192 & 0.557$\pm$0.184 & 0.545$\pm$0.186 \\
			Composed & 0.571$\pm$0.261 & 0.600$\pm$0.237 & 0.708$\pm$0.210 & 0.732$\pm$0.188 & 0.623$\pm$0.159 & 0.496$\pm$0.219 & 0.626$\pm$0.172 & 0.584$\pm$0.179 & 0.554$\pm$0.209 \\
			MixUp \cite{zhang2017mixup} & 0.586$\pm$0.239 & 0.606$\pm$0.226 & 0.692$\pm$0.238 & 0.734$\pm$0.175 & 0.621$\pm$0.165 & 0.495$\pm$0.231 & 0.620$\pm$0.176 & 0.573$\pm$0.180 & 0.554$\pm$0.188 \\
			CutMix \cite{yun2019cutmix} & \textcolor{blue}{\textbf{0.612$\pm$0.224}} & \textcolor{blue}{\textbf{0.619$\pm$0.223}} & \textcolor{blue}{\textbf{0.713$\pm$0.199}} & \textcolor{red}{\textbf{0.742$\pm$0.174}} & 0.620$\pm$0.171 & \textcolor{blue}{\textbf{0.515$\pm$0.228}} & 0.628$\pm$0.184 & \textcolor{blue}{\textbf{0.586$\pm$0.175}} & \textcolor{blue}{\textbf{0.570$\pm$0.172}} \\
			\midrule
			TBI-GAN & \textcolor{red}{\textbf{0.675$\pm$0.207}} & \textcolor{red}{\textbf{0.663$\pm$0.229}} & \textcolor{red}{\textbf{0.717$\pm$0.229}} & \textcolor{blue}{\textbf{0.736$\pm$0.204}} & \textcolor{red}{\textbf{0.681$\pm$0.177}} & \textcolor{red}{\textbf{0.526$\pm$0.202}} & \textcolor{red}{\textbf{0.654$\pm$0.205}} & \textcolor{red}{\textbf{0.605$\pm$0.192}} & \textcolor{red}{\textbf{0.592$\pm$0.184}} \\
			\toprule
			  & CRP & CLA & CLP & MCR & MCL & IPL & IPR & B & average  \\
			\midrule
			Baseline & 0.582$\pm$0.136 & 0.545$\pm$0.146 & 0.594$\pm$0.112 & 0.665$\pm$0.102 & 0.643$\pm$0.136 & 0.537$\pm$0.202 & 0.498$\pm$0.242 & 0.873$\pm$0.121 & 0.610$\pm$0.208 \\
			Spatial & 0.603$\pm$0.126 & 0.556$\pm$0.168 & 0.634$\pm$0.079 & 0.658$\pm$0.164 & 0.650$\pm$0.119 & 0.536$\pm$0.182 & 0.539$\pm$0.238 & 0.886$\pm$0.092 & 0.620$\pm$0.204 \\
			Intensity & 0.588$\pm$0.125 & 0.542$\pm$0.163 & 0.602$\pm$0.117 & 0.666$\pm$0.117 & 0.634$\pm$0.158 & 0.539$\pm$0.167 & 0.512$\pm$0.244 & 0.878$\pm$0.108 & 0.612$\pm$0.206 \\
			Composed & 0.602$\pm$0.118 & 0.551$\pm$0.170 & 0.626$\pm$0.080 & 0.664$\pm$0.172 & 0.651$\pm$0.133 & 0.521$\pm$0.195 & 0.533$\pm$0.253 & 0.886$\pm$0.109 & 0.620$\pm$0.208 \\
			MixUp \cite{zhang2017mixup} & 0.605$\pm$0.116 & 0.558$\pm$0.155 & 0.614$\pm$0.110 & 0.663$\pm$0.124 & 0.646$\pm$0.134 & 0.556$\pm$0.165 & 0.551$\pm$0.232 & 0.882$\pm$0.064 & 0.621$\pm$0.198 \\
			CutMix \cite{yun2019cutmix} & \textcolor{blue}{\textbf{0.616$\pm$0.102}} & \textcolor{blue}{\textbf{0.574$\pm$0.160}} & \textcolor{blue}{\textbf{0.635$\pm$0.088}} & \textcolor{blue}{\textbf{0.677$\pm$0.109}} & \textcolor{blue}{\textbf{0.669$\pm$0.119}} & \textcolor{blue}{\textbf{0.572$\pm$0.164}} & \textcolor{red}{\textbf{0.562$\pm$0.224}} & \textcolor{red}{\textbf{0.889$\pm$0.073}} & \textcolor{blue}{\textbf{0.635$\pm$0.191}} \\
			\midrule
			TBI-GAN & \textcolor{red}{\textbf{0.642$\pm$0.128}} & \textcolor{red}{\textbf{0.610$\pm$0.142}} & \textcolor{red}{\textbf{0.652$\pm$0.120}} & \textcolor{red}{\textbf{0.721$\pm$0.084}} & \textcolor{red}{\textbf{0.679$\pm$0.123}} & \textcolor{red}{\textbf{0.600$\pm$0.209}} & \textcolor{blue}{\textbf{0.552$\pm$0.254}} & \textcolor{blue}{\textbf{0.889$\pm$0.141}} & \textcolor{red}{\textbf{0.658$\pm$0.201}} \\
			\bottomrule
	\end{tabular}}
\end{table*}
\subsection{Comparison with Alternative Data Augmentation Techniques in Segmentation}
\subsubsection{Baseline Data Augmentations}
We validate the data augmentation capacity of the proposed TBI-GAN by comparing the proposed method with the following common data augmentation techniques: 1) spatial augmentation including random affine transform and random grid distortion; 2) intensity transform including random brightness and random contrast; 3) composed augmentation consisting of both the spatial augmentation and intensity augmentation above; 4) MixUp \cite{zhang2017mixup} which mix two images by linear addition and 5) CutMix \cite{yun2019cutmix} which cuts and pastes image patches among different images. 1) - 3) are the most common data augmentation practices used in medical segmentation, while 4) and 5) are the state-of-the-art data augmentation techniques in recent years.
\subsubsection{Implementation Details}
We implement 2D and 3D U-Net with PyTorch 1.9.0 framework on Ubuntu servers with NVIDIA RTX 3090 GPUs. PyTorch native mixed-precision training is adopted as well to accelerate training. We have synthesized 200 TBI MR scans with paired brain region labels based on 35 normal brain templates. 5-fold cross-validation is adopted to evaluate the performance of the segmentation network. Note that during cross-validation, the validation set is hidden to both the TBI-GAN and the segmentation network, which avoids any potential data leakage. 

All the segmentation networks are optimized with the Novograd optimizer \cite{ginsburg2019stochastic} with an initial learning rate of $10^{-3}$ decayed following the cosine annealing strategy. For 2D segmentation, we directly perform segmentation on the generated axial slices, and the batch size is set to 48. Batch normalization \cite{ioffe2015batch} and ReLU activation functions are utilized in the 2D U-Net. For 3D segmentation, we crop random patches from the volumes during training and use a patch-based inference pipeline during testing. The patch size is set to $128 \times 128 \times 128$ and the batch size is set to 1. We utilize instance normalization \cite{ulyanov2016instance} and leaky ReLU activation functions in the 3D U-Net, which have proved to be superior to batch normalization and ReLU in many 3D brain segmentation tasks. We train the segmentation networks for 100 epochs in both 2D and 3D tasks.
\subsubsection{Quantitative Analysis}
\begin{figure*}[ht]
	\centering
	\includegraphics[width=0.85\textwidth]{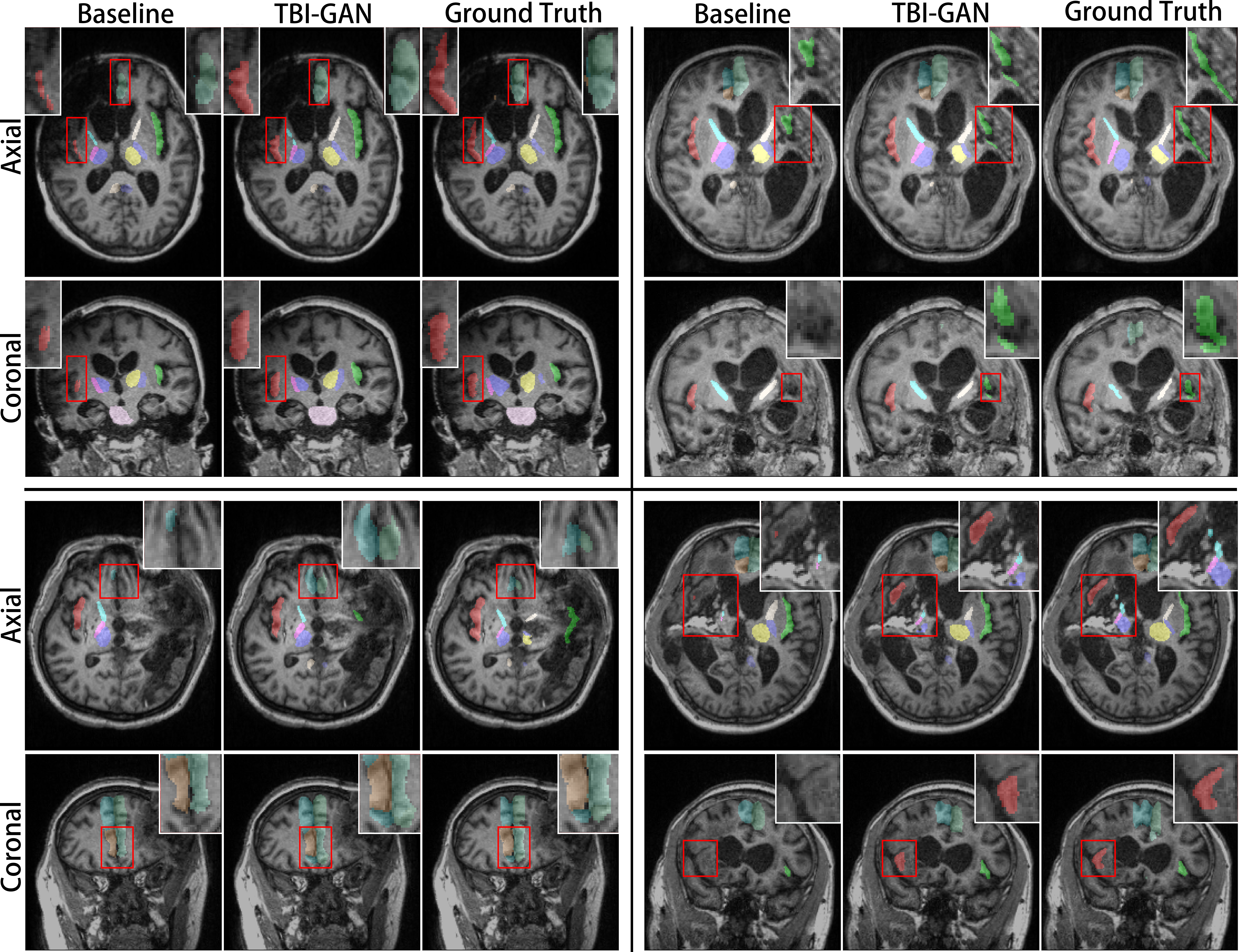} 
	\caption{Visualization of 3D segmentation results.}
	\label{fig:seg_vis}
\end{figure*}
Detailed comparisons among different augmentation techniques in both 2D and 3D segmentation tasks are shown in Table \ref{table:2dquantitative} and Table \ref{table:3dquantitative}, respectively. From Table \ref{table:2dquantitative} and Table \ref{table:3dquantitative}, it can be observed that the proposed TBI-GAN yields the best data augmentation performance in both 2D and 3D segmentation tasks, improving the brain segmentation performance by a large margin compared with the baseline method. In 2D segmentation tasks, the proposed TBI-GAN achieves the highest or the second-highest Dice score on 13 brain regions out of the total 17 brain regions. For conventional data augmentation techniques, the composed augmentation of both spatial and intensity transformations yields moderate data augmentation performance. For 3D segmentation, the proposed TBI-GAN also has the highest or the second-highest Dice score on all of the 17 brain regions. For conventional augmentation techniques, CutMix achieves the second-highest overall Dice score among the 17 brain regions. Although MixUp and CutMix have been quite popular in natural image segmentation, the performance boost brought by these methods is limited in both 2D and 3D segmentation tasks. This is because these methods cannot synthesize realistic images, and are unable to simulate the natural variance of human brains. Spatial transformation can bring about moderate performance boot in both 2D and 3D tasks, while intensity transformation is useful only when combined with spatial transformation. Despite the improvement provided by these conventional data augmentation techniques, the performance elevation is limited as these methods can not mimic the natural structures of traumatic brains. Since the proposed TBI-GAN can synthesize high-quality traumatic brain images and labels, the performance gain in both 2D and 3D segmentation tasks is significant. 
\subsubsection{Qualitative Analysis}
A visual comparison of different augmentation techniques is shown in Fig. \ref{fig:seg_vis}. Compared with the baseline model, the proposed TBI-GAN also improves the segmentation performance significantly. When segmenting certain brain regions, such as Insula-R and Capsule-R-Ant, the baseline model is rather unstable and tends to produce false-negative easily. This is because such brain regions tend to be affected by the lesions caused by traumatic injuries. It can also be concluded that the proposed TBI-GAN can boost the augmented segmentation model in dealing with these brain regions and reduce mis-segmentation significantly, due to its effectiveness in generating the realistic-looking TBI image with correct labels.


\section{Conclusion}
\label{sec:conclusion}
In this paper, we develop a novel generative adversarial network, TBI-GAN, for the data augmentation of brain region segmentation in TBI MR scans. The proposed TBI-GAN can inpaint both the MR image and brain region label maps based on normal brain templates, which is suitable for the data augmentation of traumatic brain segmentation. The segmentation network augmented by our proposed method yields the best performance in both 2D and 3D brain region segmentation in TBI MR scans. The experimental results reveal that the proposed method has great potential to be applied in the traumatic brain segmentation tasks and benefit the clinical healthcare of TBI patients. 

In future works, we plan to recruit more TBI patients with severe brain images and collect their MR images to further validate the effectiveness of our brain segmentation model aided with TBI-GAN. We will also conduct further investigations on applying the proposed TBI-GAN method to other brain images with symptoms of large lesions, such as gliomas and cerebral aneurysm, which can also help boost their brain region segmentation works.

\bibliography{references}

\end{document}